\pdfoutput=1
\documentclass[12pt]{iopart}

\usepackage{iopams}
\usepackage{dcolumn}
\usepackage{graphicx}
\usepackage{bm}

\newcommand{\BFA}{BaFe$_{2}$As$_{2}$}
\newcommand{\SFA}{SrFe$_{2}$As$_{2}$}
\newcommand{\EFA}{EuFe$_{2}$As$_{2}$}
\newcommand{\KFA}{KFe$_{2}$As$_{2}$}
\newcommand{\BKFA}{(Ba$_{1-x}$K$_x$)Fe$_{2}$As$_{2}$}
\newcommand{\BKSC}{(Ba$_{0.6}$K$_{0.4}$)Fe$_{2}$As$_{2}$}
\newcommand{\BKld}{(Ba$_{0.9}$K$_{0.1}$)Fe$_{2}$As$_{2}$}
\newcommand{\BKmd}{(Ba$_{0.8}$K$_{0.2}$)Fe$_{2}$As$_{2}$}
\newcommand{\BKhd}{(Ba$_{0.7}$K$_{0.3}$)Fe$_{2}$As$_{2}$}
\newcommand{\TCS}{ThCr$_{2}$Si$_{2}$}

\newcommand{\MB}{$^{57}$Fe-Mö\"ossbauer}

\begin{document}



\title[Underdoped \BKFA]{Competition of magnetism and superconductivity in underdoped \BKFA}

\author{Marianne Rotter$^1$, Marcus Tegel$^1$, Inga Schellenberg$^2$ , Falko M. Schappacher$^2$, Rainer P\"ottgen$^2$,
Joachim Deisenhofer$^3$, Axel G\"{u}nther$^3$, Florian Schrettle$^3$, Alois Loidl$^3$, and Dirk Johrendt$^1$}

\address{$^1$Department Chemie und Biochemie der Ludwig-Maximilians-Universit\"{a}t M\"{u}nchen, Butenandtstr- 5-13 (Haus D), 81377 M\"{u}nchen, Germany}
\address{$^2$Institut f\"{u}r Anorganische und Analytische Chemie, Universit\"{a}t M\"{u}nster, Corrensstrasse 30,
D-48149 M\"{u}nster, Germany}
\address{$^3$Experimentalphysik V, Center for Electronic Correlations and Magnetism,
Institute for Physics, Augsburg University, D-86135 Augsburg, Germany}

\ead{johrendt@lmu.de}

\begin{abstract}

Polycrystalline samples of underdoped \BKFA~($x \leq$ 0.4) were synthesized and studied by x-ray powder diffraction, magnetic susceptibility, specific heat and \MB-spectroscopy. The structural phase transition from tetragonal to orthorhombic lattice symmetry shifts towards lower temperatures, becomes less pronounced at $x$ = 0.1-0.2 and is no longer present at $x$ = 0.3. Bulk superconductivity is observed in all samples except \BKld~by resistivity and magnetic susceptibility measurements. Specific heat data show a broad SDW phase transition in \BKld, which is hardly discernible in \BKmd. No SDW anomaly is found in the specific heat of optimally doped \BKSC, where $C$ changes by 0.1 J/K at $T_c$ = 37.3 K. \MB-spectra show full magnetic hyperfine field splitting, indicative of antiferromagnetic ordering at 4.2 K in samples with $x$ = 0-0.2, but zero magnetic hyperfine field in samples with $x$ = 0.3. The spectra of \BKld~and \BKmd~in the phase transition regions are temperature-dependent superpositions of magnetic and non-magnetic components, caused by inhomogeneous potassium distribution. Our results suggest the co-existence of AF magnetic ordering and superconductivity without mesoscopic phase separation in the underdoped region and show unambiguously homogeneous superconducting phases close to optimal doping. This is in contrast to recently reported results about single crystal \BKFA.

\end{abstract}




\submitto{\NJP}

\maketitle

\section{Introduction}

The discovery of superconductivity (SC) in iron arsenides \cite{Hosono-2008, BKFA} with transition temperatures ($T_c$) up to 55 K \cite{Sm-TC55} has attracted an enormous interest in the scientific community \cite{Angew-2008}. Besides the outstanding physical properties of this new class of superconducting materials, scientists found fresh hope that iron arsenides may help to finally solve the mystery of high-$T_c$ superconductivity. But prior to this long-term objective, many fundamental issues of the iron arsenides need to be clarified. Among them, the structural and magnetic phase diagrams with respect to doping, reflecting the interplay between superconductivity and magnetism, are discussed controversially.

In both the LaFeAsO (1111) and \BFA~(122) families, superconductivity evolves from poor metallic parent compounds with quasi two-dimensional tetragonal crystal structures, which are subject to orthorhombic lattice distortions below certain temperatures ($T_o$). Static long-range antiferromagnetic (AF) ordering emerges with N\'{e}el temperatures ($T_N$) well below $T_o$ in LaFeAsO \cite{Cruz-Neutrons}, but very close to $T_o$ in \BFA~\cite{BFA}. The structural and magnetic transitions of the parent compounds are strongly affected by doping of the FeAs layers either with electrons or holes, and superconductivity appears at certain doping levels. For the underdoped phases in the transition zone, it has been reported that SC and AF ordering is either separated or co-existing. Also the overlap of the orthorhombic distortion with the SC and AF areas in the phase diagrams are still not clear, neither in the 1111 nor in the 122 systems.

The first phase diagram of LaFeAsO$_{1-x}$F$_x$, constructed by $\mu$SR data, showed a sharp-cut vertical border between the SC and the orthorhombic AF phases at $x$ = 0.045 \cite{Luetkens-2008}. But neutron diffraction experiments showed, that although the magnetic ordering vanishes around $x \approx$ 0.04, the orthorhombic lattice still exists at least to $x$ = 0.05, where superconductivity has already emerged \cite{Huang-2008}. This is in line with the recently published neutron study of CeFeAsO$_{1-x}$F$_x$, where AF ordering disappears exactly before SC emerges, but the orthorhombic lattice persists extensively into the SC dome up to $x \approx$ 0.1 \cite{Zhao-2008}. Similar results have been reported for SmFeAsO$_{1-x}$F$_x$ from $\mu$SR experiments \cite{Drew-2008} and structural investigations using synchrotron radiation \cite{Margadonna-2008}. Thus at the moment all signs are that in the case of the 1111-family, static AF order is completely suppressed before SC emerges, but the orthorhombic lattice co-exists with superconductivity and the temperature difference between $T_o$ and $T_N$ increases with the doping level. This behavior of the 1111-superconductors is strongly reminiscent to the monolayer high-$T_c$ cuprates. For instance in La$_{2-x}$Sr$_x$CuO$_4$, the AF order is well separated from the SC state, but the orthorhombic phase exists largely in the superconducting dome \cite{Keimer-1992}.

In the 122-family,  co-existence of the orthorhombic structure with SC has been first published for \BKFA~up to $x \approx$ 0.2 ($T_c \approx$ 26 K) by X-ray powder diffraction \cite{Rotter-Angewandte}. Following neutron diffraction experiments also showed orthorhombic symmetry and long-range AF ordering co-existing up to $x$ = 0.3 ($T_c <$ 15 K) \cite{Chen-BKFA}. The different shapes of the superconducting domes $T_c(x)$ of \BKFA~may be due to different synthesis conditions. However, the $x$ values in Ref. \cite{Rotter-Angewandte} are determined from X-ray data by Rietveld refinements, whereas only the nominal compositions are given in Ref. \cite{Chen-BKFA}. Since diffraction methods provide the mean structural information on a rather long spatial scale, short-range phase inhomogeneities are averaged. Thus one may understand the observed co-existence of SC with AF ordering in \BKFA~by phase-separation in magnetic non-superconducting and non-magnetic superconducting mesoscopic domains. Local probes like $\mu$SR and \MB-spectroscopy can provide more accurate information. Recently, three reports about $\mu$SR experiments, each conducted with almost optimally doped superconducting \BKFA~single crystals, concluded consistently phase separations into SC and AF domains. The non-magnetic superconducting volume fractions were found to be $\approx$ 30\% \cite{Aczel-2008}, 40\% \cite{Goko-2008}, and 25\% \cite{Keimer-2008}. In the latter report, the lateral scale of of the inhomogeneities were estimated to 65$\pm$10 nm by magnetic force microscopy (MFM) imaging. However, the onset of AF ordering in the superconducting crystals were detected at $\approx$ 70-80 K irrespective of different doping levels.

In the present paper, we report on a detailed study of the structural and magnetic transitions of polycrystalline underdoped \BKFA~($x \leq$ 0.4). The samples were characterized by magnetic susceptibility and specific heat measurements. The crystal structures and chemical compositions were determined by Rietveld refinements of x-ray powder patterns. Detailed temperature-dependent \MB-spectra were recorded in order to detect the evolution of magnetic ordering on a local spatial scale.

\section{Experimental}

\subsection{Sample preparation}

Samples of \BKFA~with $x$ = 0.1, 0.2 and 0.3 were prepared by heating stoichiometric mixtures of the elements (all purities $>$ 99.9 \%) in alumina crucibles enclosed in silica tubes under an atmosphere of purified argon. In order to minimize the loss of potassium by evaporation at elevated temperatures, the gas volume in the crucibles was reduced by alumina inlays. The mixtures were heated slowly (50 K/h) to 873 K, kept at this temperature for 15 h and cooled down to room temperature. The reaction products were homogenized in an agate mortar and annealed at 923 K for 15 h. After cooling, the samples were homogenized again, pressed into pellets and sintered at 1023 K for 15 h. The obtained black crystalline powders are stable in air for weeks. The Ba:K ratios were checked by EDX and chemical analysis (ICP-AAS), which resulted in the nominal composition within 5\%.

\subsection{X-ray structure determination}

Phase purity was checked by X-ray powder diffraction using a Huber G670 Guinier imaging plate diffractometer (Cu-$K_{\alpha_{1}}$ radiation, Ge-111 monochromator), equipped with a closed-cycle He-cryostat. Rietveld refinements of all diffractograms were performed with the TOPAS package \cite{TOPAS} using the fundamental parameters approach as reflection profiles (convolution of appropriate source emission profiles with equatorial and axial instrument contributions as well as crystallite microstructure effects). In order to obtain crystal structures inclusive of the Ba:K ratios, all profile contributions were refined freely, but in order to obtain accurate lattice parameter changes, all profile contributions were refined at room temperature and held constant for all other temperatures (except for the lorentzian strain contribution). All diffractograms were measured without an internal standard, so the absolute lattice parameters might be slightly offset. In all cases, an empirical 2$\theta$-dependent intensity correction for different absorption lengths arising from the Guinier geometry setup was applied.

\subsection{Magnetic susceptibility and specific heat}

The magnetic properties were studied using a commercial SQUID magnetometer (Quantum Design MPMS-5) with external magnetic fields up to 50~kOe. The heat capacity was measured in a Quantum Design Physical Properties Measurement System for temperatures from 2~K to 300~K\@.

\subsection{M\"ossbauer spectroscopy}

A $^{57}$Co/Rh source was available for the \MB~spectroscopy investigations. The samples were placed in thin-walled PVC containers at a thickness of about 10 mg Fe/cm$^2$. The measurements were run in the usual transmission geometry in the temperature range from room temperature to 4.2 K. The source was kept at room temperature. The total counting times per spectrum ranged between 5 h and 1 day.

\section{Results and Discussion}

\subsection{Crystal structures and phase transition}

\begin{figure}[h]
\center{
\includegraphics[width=0.6\textwidth]{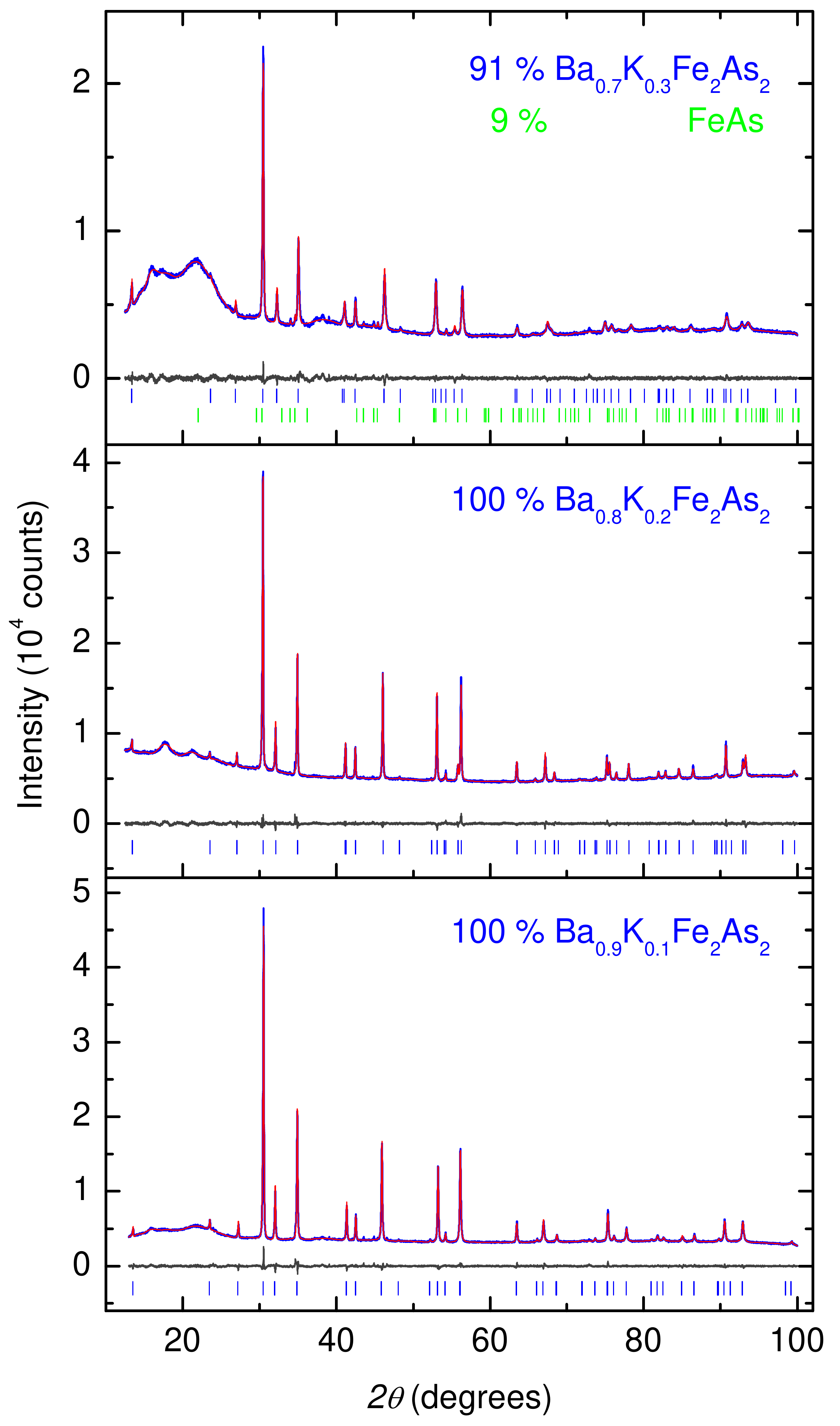}
\caption{\label{fig:xray} (Color online) X-ray powder patterns of \BKFA~(Crystallographic $x$ = 0.13, 0.20, 0.24) with Rietveld profile fits and difference lines.}
}
\end{figure}

\begin{figure}[h]
\center{
\includegraphics[width=0.7\textwidth]{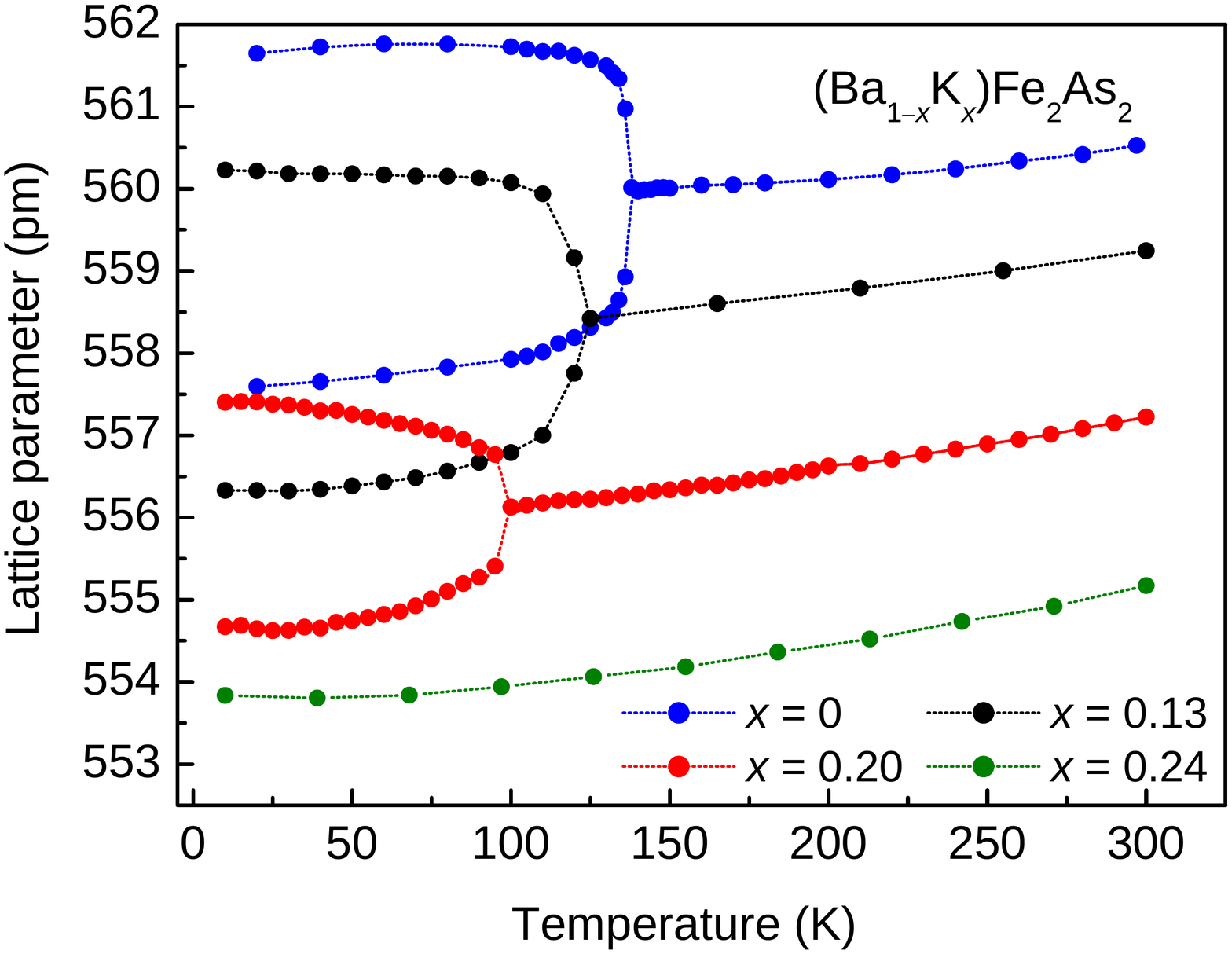}
\caption{\label{fig:Lattices} (Color online) Lattice parameters of \BKFA~(Crystallographic $x$ = 0, 0.13, 0.20, 0.24). The tetragonal parameters are multiplied by $\sqrt{2}$ for comparison.}
}
\end{figure}

The crystal structures were determined by x-ray powder diffraction. Fig.~\ref{fig:xray} shows the x-ray powder patters of \BKFA~($x$ = 0.1, 0.2, 0.3) at 300 K with Rietveld-fits and the difference lines. Crystallographic data and selected bond lengths and angles at 300 K and 10 K, respectively, are compiled in Table~\ref{tab:Structures}. The temperature dependencies of the $a$ and $b$ lattice parameters are shown in Fig.~\ref{fig:Lattices}. In line with Ref.~\cite{Rotter-Angewandte}, the parameter $a$ of the tetragonal phase decreases with the doping level $x$, while $c$ increases (not shown). The tetragonal-to-orthorhombic phase transition is strongly affected by the potassium content. The transition temperatures $T_o$ decreases to $\approx$ 100 K at x = 0.2 and is no longer visible at $x$ = 0.3. Also the magnitude of the distortion, expressed by the differences between $a$ and $b$ at 10 K, decreases from 0.73\% ($x$ = 0) to 0.70\% ($x$ = 0.1) to 0.49\% ($x$ = 0.2). Thus with increasing potassium doping levels, the structural transition of \BFA~is shifted towards lower temperatures and also less pronounced. It is no longer present at $x$ = 0.3 (see Fig.~\ref{fig:Lattices}).

\subsection{$dc$ resistivity}

\begin{figure}[h]
\center{
\includegraphics[width=0.7\textwidth]{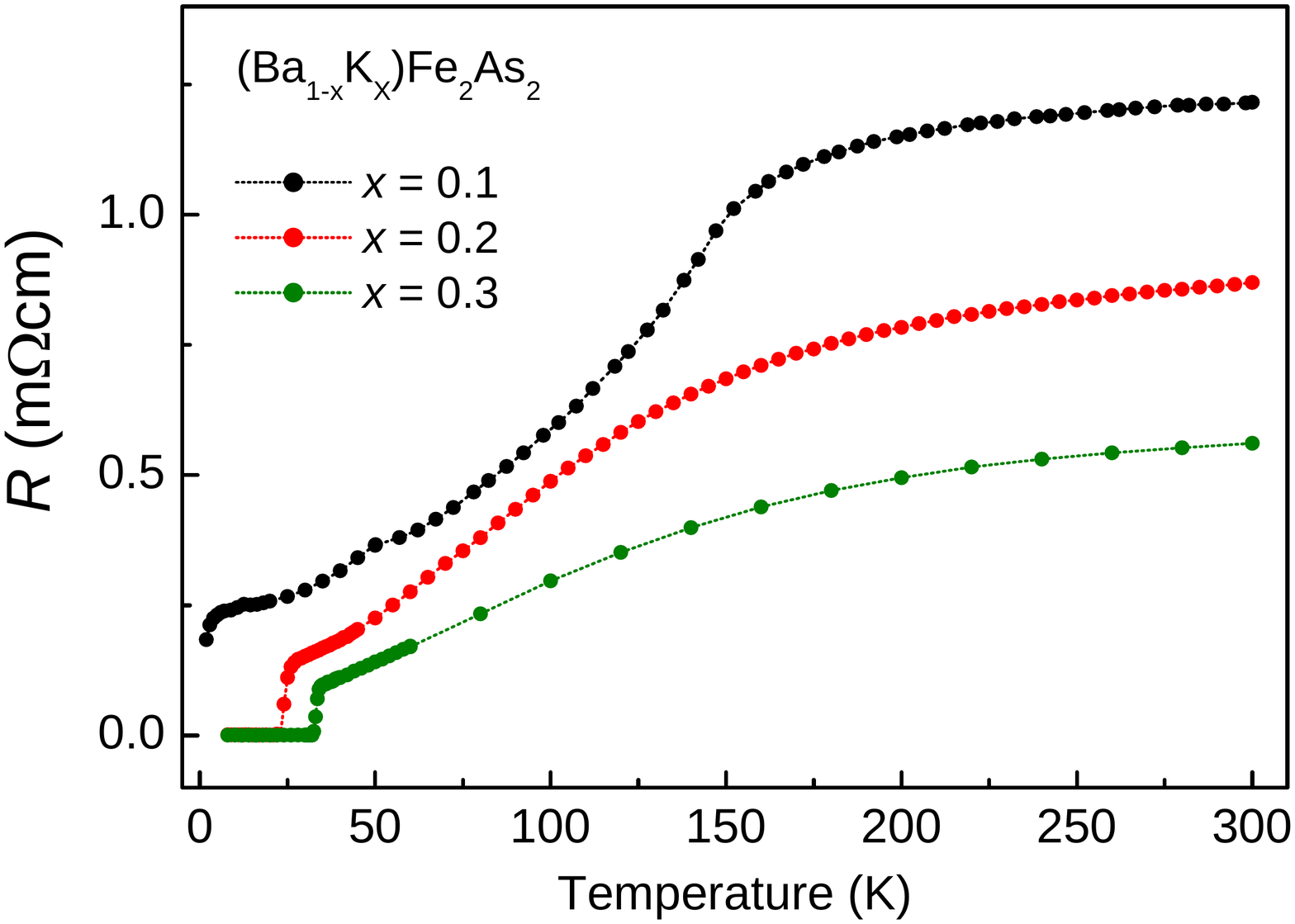}
\caption{\label{fig:Resistance} (Color online) Temperature dependence of the $dc$ resistivity of \BKFA~($x$ = 0.1, 0.2, 0.3)}.
}
\end{figure}

The temperature dependence of the $dc$ resistivity of \BKFA~is shown in Fig.~\ref{fig:Resistance}. At the lowest doping concentration ($x$ = 0.1), the typical SDW anomaly is still visible, but shifted towards lower temperatures and less pronounced than in undoped \BFA~\cite{BFA}. We observe a drop of the resistance below 3 K, associated with a superconducting transition, even though zero resistance could not be reached at 1.8 K. The curvature of the resistivity of \BKmd~is still reminiscent to a SDW anomaly, but smeared over a larger temperature range between $\approx$ 120 K and 70 K. The superconducting transition at 24 K is rather broad ($\approx$ 4 K), but zero resistance is clearly observed at 23 K. At the higher doping level $x$ = 0.3, the resistivity shows no more indications of the SDW anomaly and superconductivity emerges at $T_c$ = 33 K.\\

\subsection{Magnetic Susceptibility}

\begin{figure}[h]
\centering
\includegraphics[width=0.7\textwidth]{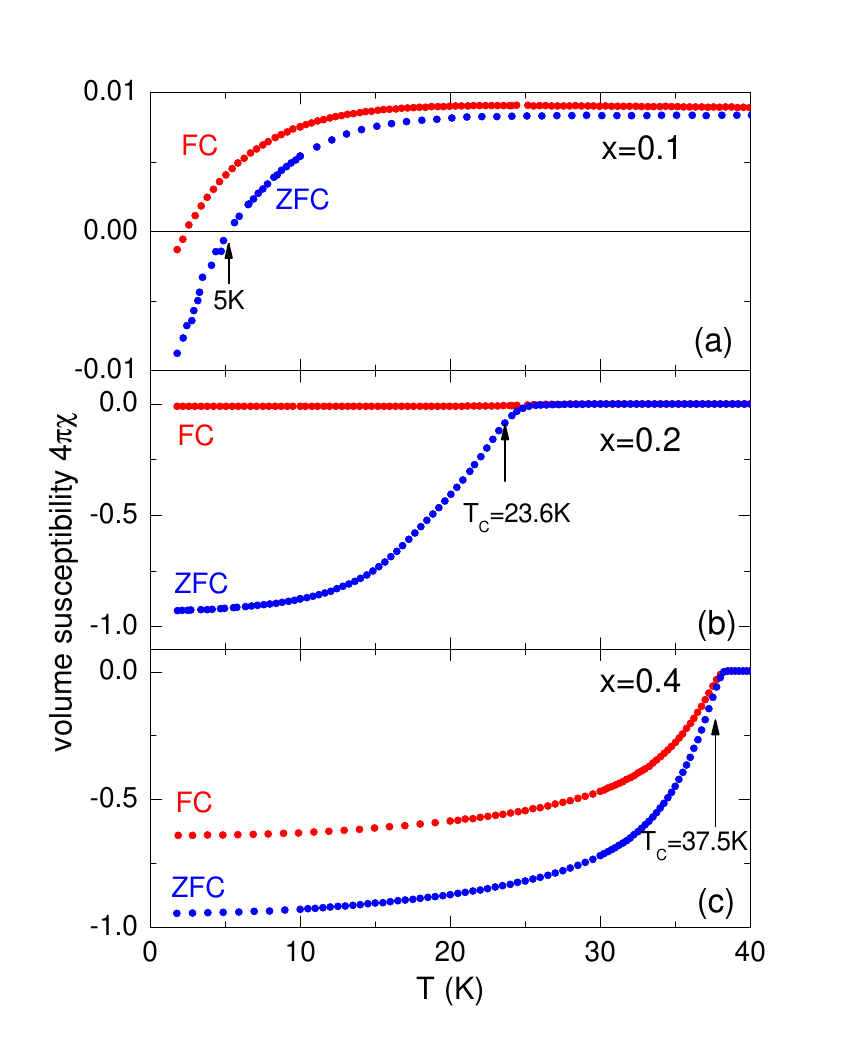}
\caption[]{\label{FC_ZFCin5Oe} (Color online) Static volume susceptibility $4\pi\chi=4\pi M/H$ for $x$ = 0.1, 0.2, and 0.4 in a magnetic field of 5~Oe for field-cooled (FC) and zero-field cooled (ZFC) cycles.}
\end{figure}

Field-cooled (FC) and zero-field cooled (ZFC) cycles of the static magnetic volume susceptibility are shown as a function of temperature in a magnetic field of 5~Oe in Fig.~\ref{FC_ZFCin5Oe}. Estimating the superconducting volume fractions for $x$=0.2 and $x$=0.4 from the ZFC value at 1.8~K as 0.93 and 0.94, respectively, bulk superconductivity is evidently present. The FC values at 1.8~K amount to 1\% and 64\%, respectively. The corresponding temperatures, where 10\% of the maximum shielding is reached, are $T_C$=23.6~K and 37.5~K for these two doping levels. For the sample with $x$=0.1 bulk superconductivity cannot be established, but the sample becomes diamagnetic below 5~K in the ZFC cycle as shown in Fig.~\ref{FC_ZFCin5Oe}(a). The positive contribution to the susceptibility may be due to ferromagnetic impurities.

\subsection{Specific heat}

\begin{figure}[h]
\centering
\includegraphics[width=0.7\textwidth]{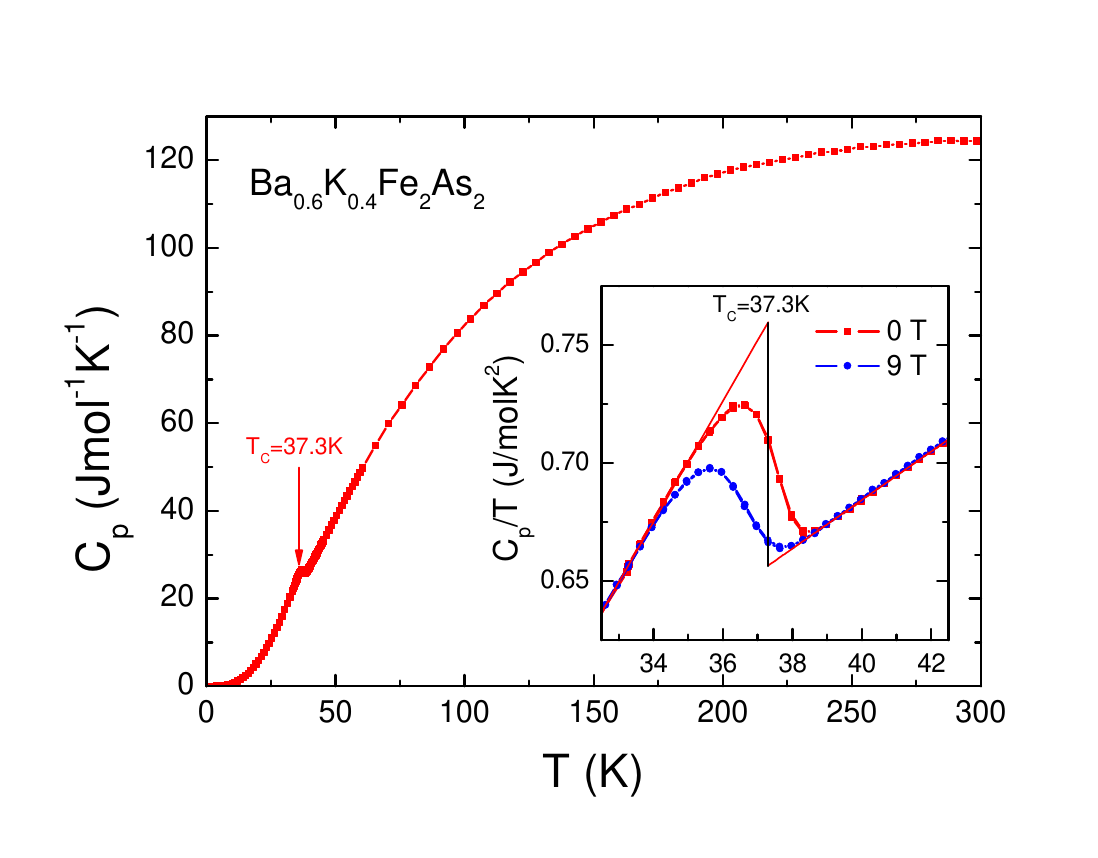}
\vspace{2mm} \caption[]{\label{C04Tc} (Color online) Specific heat of Ba$_{0.6}$K$_{0.4}$Fe$_2$As$_2$. Inset: Comparison of C/T vs. temperature in magnetic fields of 0 and 9T. Lines are to guide the eyes.}
\end{figure}

In Fig.~\ref{C04Tc} we show the temperature dependence of the specific heat for the optimally doped sample with $x$ = 0.4. In the inset $C_p/T$ is plotted as a function of temperature for zero magnetic field and a field of 9~T. Both curves are on top of each other above and below the superconducting transition. $T_C$ in zero field is 37.3~K by the entropy conserving construction shown in the inset of Fig.~\ref{C04Tc}. The change of the specific heat at the transition is estimated as $\Delta C|_{T_C}$=0.1~J/molK, which is in good agreement with other reported values \cite{Ni-2008,Welp-2008,Mu-2008}. $T_C$ is shifted by only 1~K in a field of 9~T reflecting the large upper critical fields, which were estimated as 70~T or even higher \cite{Wang-2008,Yuan-2008,Welp-2008}.

\begin{figure}[hbt]
\centering
\includegraphics[width=0.7\textwidth,clip]{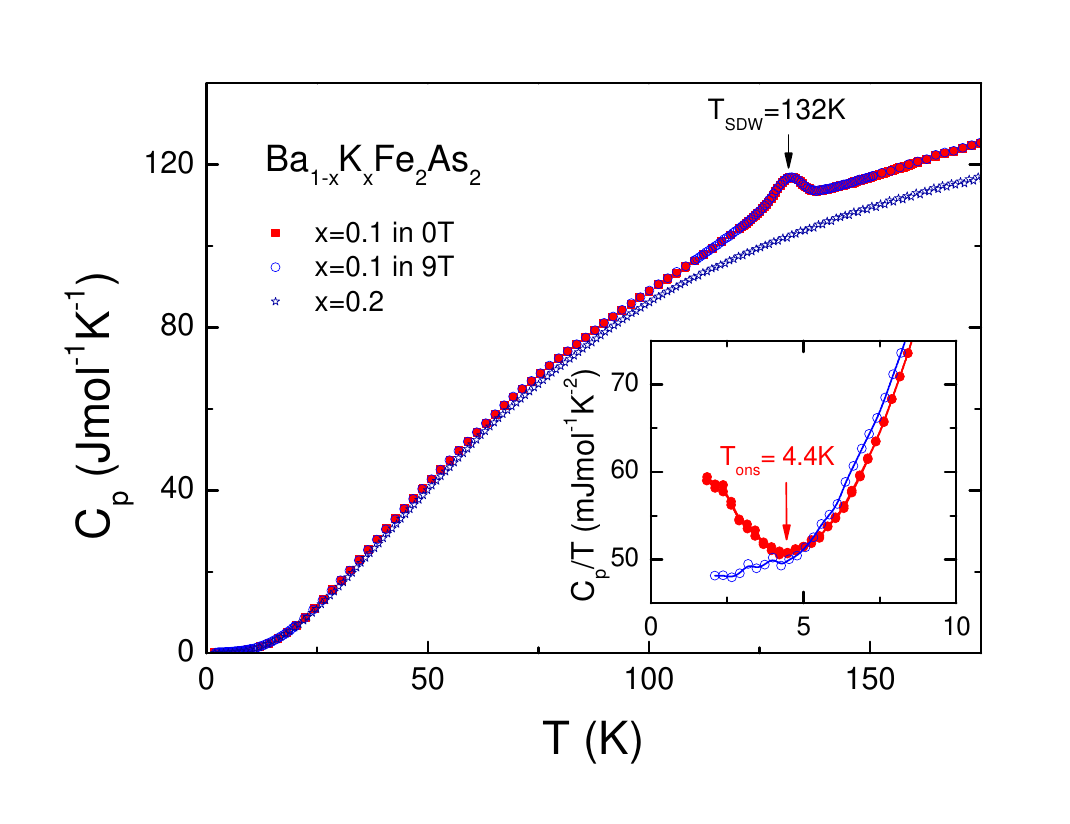}
\vspace{2mm} \caption[]{\label{C01Tc} (Color online) Comparison of C vs. $T$ for $x$=0.1 and $x$=0.2. Inset: Comparison of C/T vs. $T$ in magnetic fields of 0 and 9T for $x$=0.1 to reveal the onset of the superconducting transition and its suppression in a magnetic field of 9~T. Solid lines are to guide the eyes.}
\end{figure}

The specific heat for the underdoped samples with $x$ = 0.1 and $x$ = 0.2 is plotted as a function of temperature in Fig.~\ref{C01Tc}. For $x$ = 0.1 one can clearly see a broad peak with a maximum at $T_{SDW}$ = 132~K, which is in agreement with the structural and magnetic transitions.

The low-temperature C/T data for the sample with $x$=0.1 reveals the onset of a phase transition at $T_{ons}$=4.4~K (inset of Fig.~\ref{C01Tc}), which is completely suppressed in a magnetic field of 9~T. This temperature is in agreement with the appearance of diamagnetism as indicated in Fig.~\ref{FC_ZFCin5Oe}(a) and the appearance of a resistivity drop and therefore can be ascribed to the onset of superconductivity. Hence, a superconducting and an antiferromagnetic transition are observable for $x$ = 0.1, suggesting a strong competition of magnetic fluctuations and superconductivity. However, one must keep in mind that bulk superconductivity could not be established from susceptibility measurements and zero resistivity is not reached at 1.8~K.

\begin{figure}[hbt]
\centering
\includegraphics[width=0.8\textwidth]{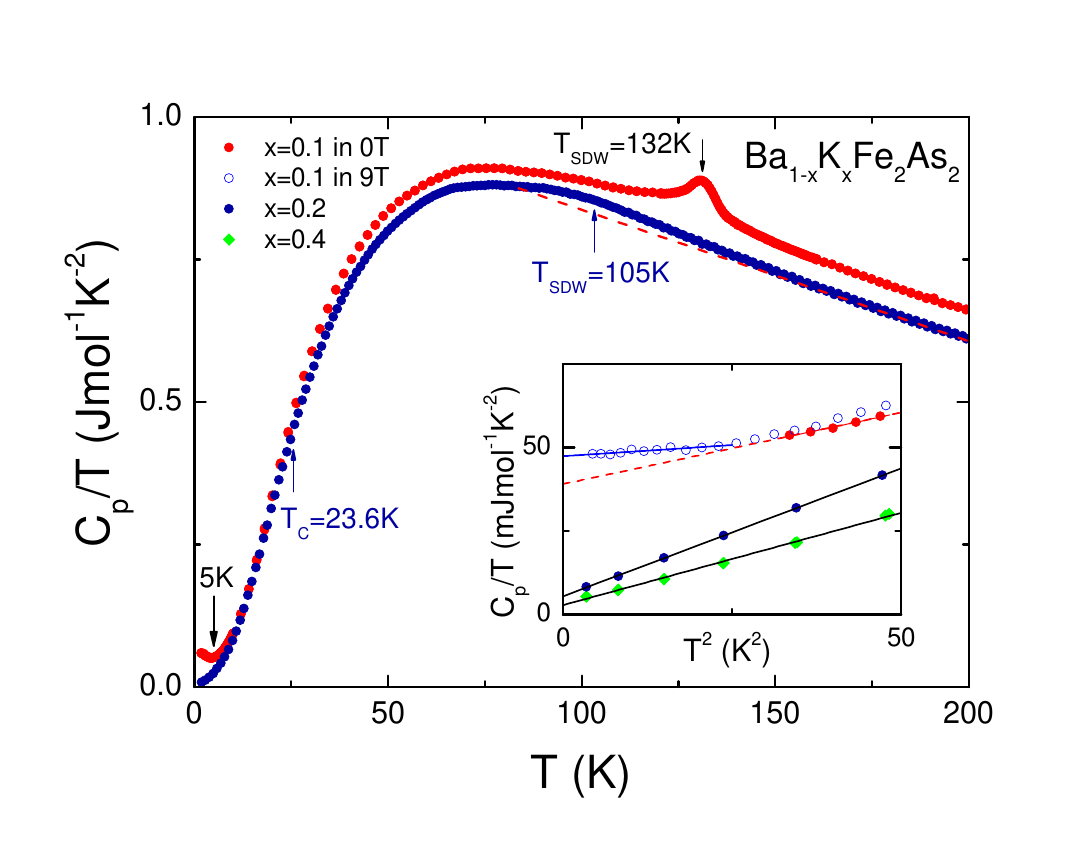}
\vspace{2mm} \caption[]{\label{Cdtlowx} (Color online) Comparison of C/T vs. $T$ for $x$=0.1 and $x$=0.2. The dashed line is a guide for the eye. Inset: Comparison of C/T vs. $T^2$ for $x$=0.1 (0 and 9T) and $x$=0.2, 0.4 at lowest temperatures. Solid and dashed lines are linear fits described in the text.}
\end{figure}

Astonishingly, we could not detect any clearly visible anomaly for the sample with $x$ = 0.2, where bulk superconductivity is evident from Fig.~\ref{FC_ZFCin5Oe}(a) below $T_C$ = 23.6~K and the structural transition occurs at $T_o$ = 105~K. In Fig.~\ref{Cdtlowx} we compare C/T vs. $T$ for these two concentrations and indicate the known transition temperatures for both compounds. While at around $T_o$ = 105~K, an extremely broadened transition region between about 80~K and 140~K may be visualized by guiding the eye with a linear extrapolation of the high-temperature behavior, a corresponding anomaly at the superconducting transition is not detectable in our data. The low-temperature data for all three concentrations $x$ = 0.1 (in 0 and 9 T) and $x$ = 0.2, 0.4 are plotted as C/T vs. $T^2$. We fitted the data for $x$ = 0.1 at 9T and for $x$ = 0.2, 0.4 in zero magnetic field with a linear behavior and extracted Sommerfeld coefficients of $\gamma$ = 47,~5.6,~2.8~mJ/molK$^2$ and Debye temperatures $\Theta_D$=418,~238,~260~K, respectively (solid lines in the inset of Fig.~\ref{Cdtlowx}. The values for $\gamma$ and $\Theta_D$ for $x$ = 0.2 and 0.4 are very close to each other and significantly lower than the values for $x$ = 0.1. This reflects the superconducting low temperature state for $x$ = 0.2 and 0.4, which is fully suppressed for $x$ = 0.1 in a magnetic field of 9~T. Note that the value $\gamma$ = 47~mJ/molK$^2$ and $\Theta_D$= 418~K may contain a magnetic-field dependent contribution, because a linear fit (dashed line in the inset of Fig.~\ref{Cdtlowx}) of the data above 5~K in zero field yield somewhat lower values of $\gamma$=39~mJ/molK$^2$ and $\Theta_D$=283~K.

\subsection{\MB~Spectroscopy}

\MB-spectroscopy is an excellent local probe for the determination of magnetic ordering in iron compounds. In comparison with other experimental methods, M\"ossbauer spectroscopy studies of the new superconductors are rather rare. So far LaFePO \cite{Tegel-LaFePO}, LaFeAsO \cite{Kitao-2008,Klauss-2008,Nowik-2008,McGuire-2008}, SrFeAsF \cite{Tegel-SrFeAsF}, \BFA \cite{BFA}, \SFA~\cite{Tegel-SFA} and \EFA~\cite{Raffius-1993} have been investigated. Since the formation of binary iron phosphide or arsenide impurities can seriously affect the property measurements, \MB~spectroscopy can also be an useful analytical tool to detect iron impurity phases. The influence of {Fe$_2$As}, FeAs, and FeAs$_2$ on the SDW transitions and the superconducting properties has systematically been studied \cite{Felner-2008}.

\begin{figure}[h]
\center{
\includegraphics[width=0.6\textwidth]{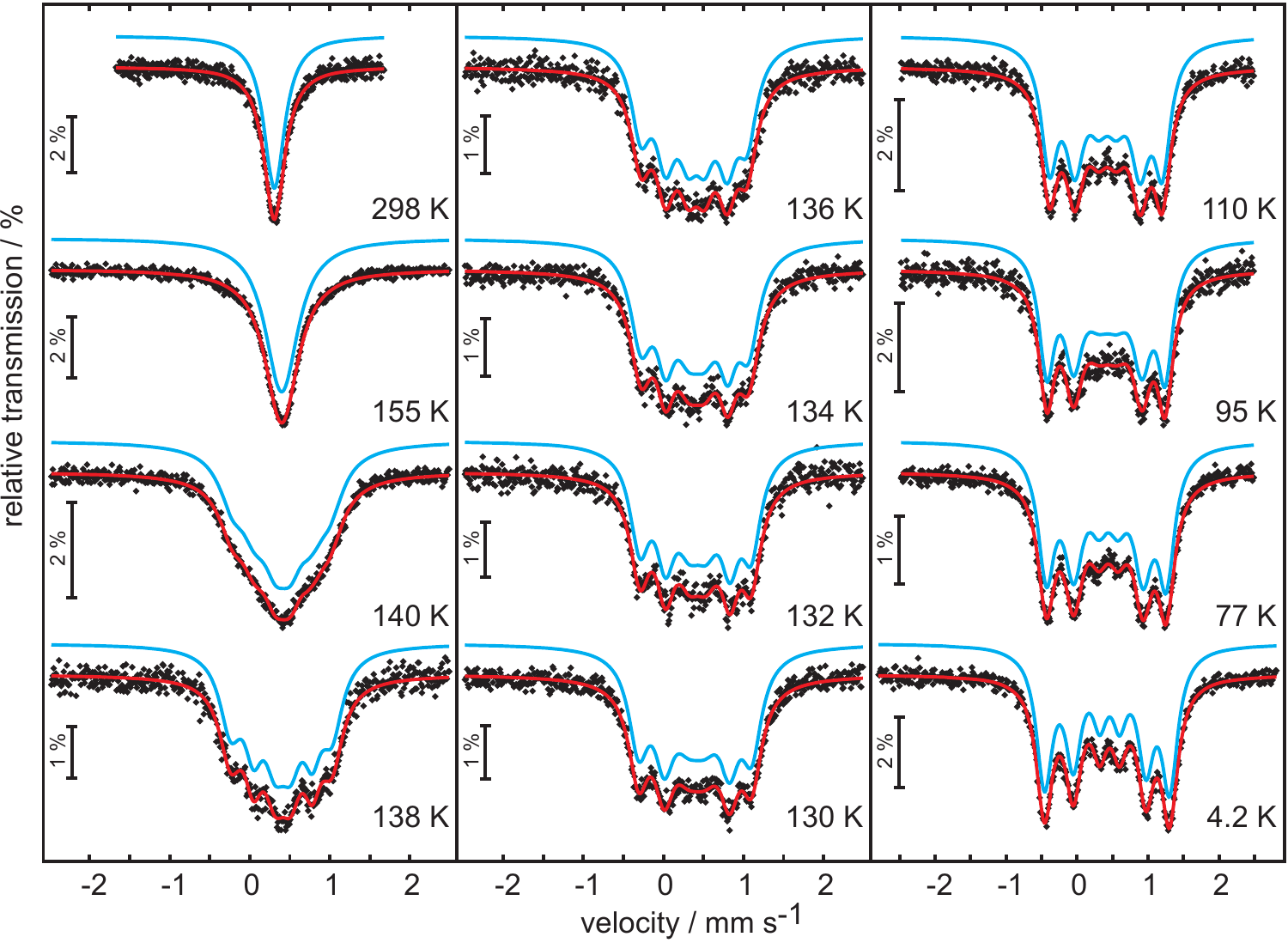}
\caption{\label{fig:MB-BFA} (Color online) \MB-spectra of \BFA.}
}
\end{figure}

First we discuss the spectra of \BFA~and \KFA. In our earlier report on the SDW anomaly of \BFA, \MB~spectra had only been recorded at 298, 77, and 4.2 K \cite{BFA}. We have now studied \BFA~over the entire temperature range. The spectra are presented in Fig.~\ref{fig:MB-BFA} together with transmission integral fits. The corresponding fitting parameters are listed in Table~\ref{tab:MB-Data}. At room temperature, the spectrum consists of a single Lorentzian line with an isomer shift of $\delta$ = 0.31(1) mms$^{-1}$. First hints at line broadening and thus short-range magnetic ordering appear already at 155 K, thus well above the structural distortion, however, with a very small magnetic hyperfine field (Table~\ref{tab:MB-Data}). When lowering the temperature below the SDW transition temperature of 138 K, we observe a strong increase of the internal hyperfine field with a saturation value of 5.47 T at 4.2 K (see Fig.~\ref{fig:MB-BFA}). This corresponds to a magnetic moment of approximately 0.4 -0.5 $\mu_B$ per iron atom. Similar behavior has been observed for \SFA~\cite{Tegel-SFA} and the quaternary iron arsenide LaFeAsO \cite{Kitao-2008,Klauss-2008,Nowik-2008,McGuire-2008}. In the latter two compounds the saturation hyperfine fields are 8.9 T (\SFA) and 5.3 T (LaFeAsO).

For \BFA~reported herein and also for \SFA~\cite{Tegel-SFA}, the magnetically split spectra can be well reproduced by one spectral component with the typical sextet of lines. This is in contrast to the ZrCuSiAs-type compounds LaFeAsO \cite{Kitao-2008} and SrFeAsF \cite{Tegel-SrFeAsF}, which both show distributions of moments in the magnetically ordered states, indicating that some spin disorder still remains. Most likely the ordering mechanisms in the ZrCuSiAs and \TCS-type materials are slightly different.\\

\begin{figure}[h]
\center{
\includegraphics[width=0.4\textwidth]{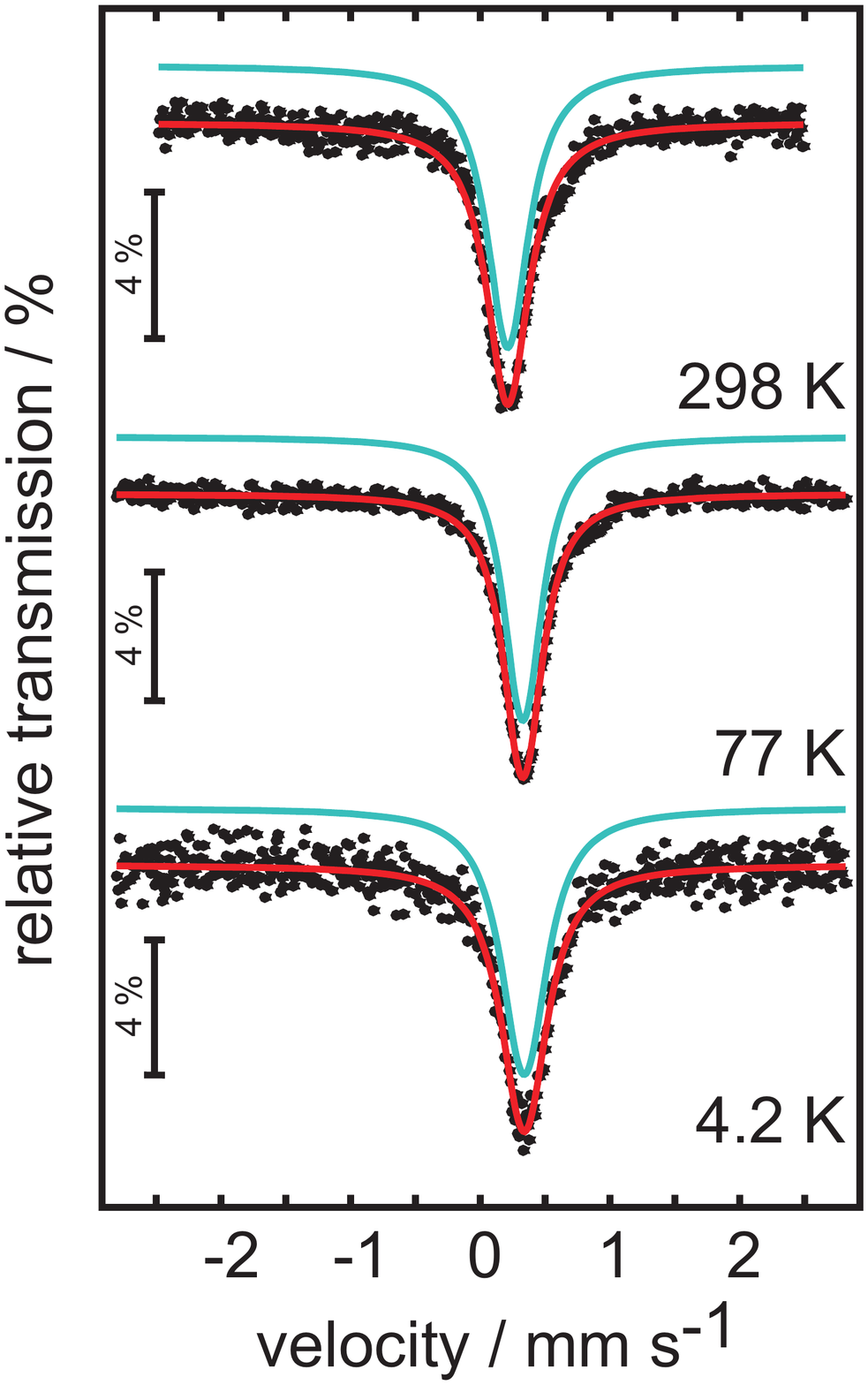}
\caption{\label{fig:MB-KFA} (Color online) \MB-spectra of \KFA.}
}
\end{figure}

The \MB-spectra of \KFA~are shown in Fig.~\ref{fig:MB-KFA}. \KFA~is a metal with a superconducting transition at $T_C$ = 3.8 K. We observe a single absorption line with an isomer shift of 0.21 mms$^{-1}$ from room temperature down to 4.2 K, which is subjected to weak quadrupole splitting (Table~\ref{tab:MB-Data}). Since potassium transfers only one electron to the [Fe$_2$As$_2$] layer, we expect a smaller electron density at the iron nuclei in \KFA, which is consistent with lower isomer shift compared to \BFA~\cite{BFA} and \SFA~\cite{Tegel-SFA}. From this it is evident that the potassium doping affects the electronic situation of the iron atoms.

\begin{figure}[h]
\center{
\includegraphics[width=0.6\textwidth]{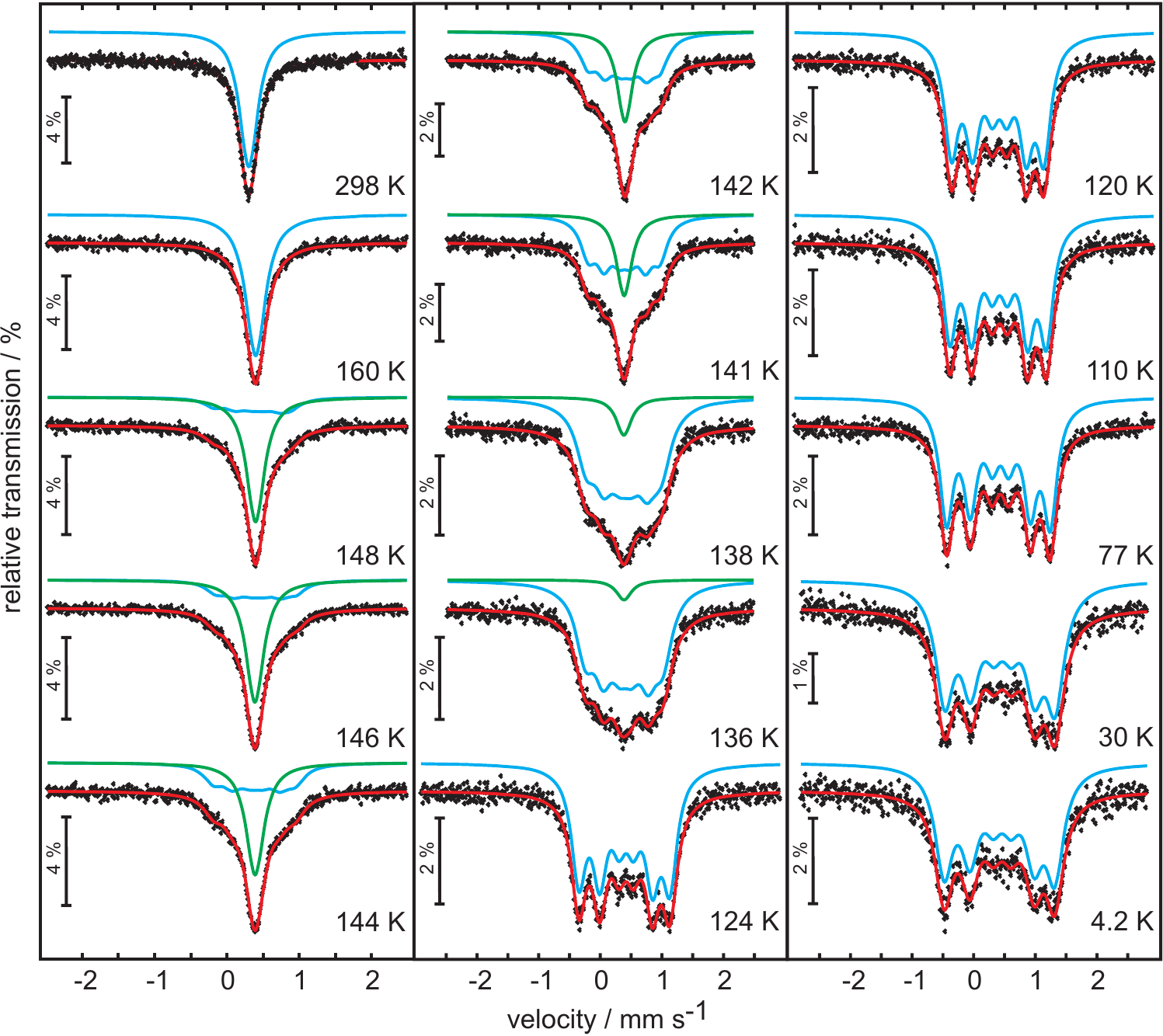}
\caption{\label{fig:MB-BKld} (Color online) \MB-spectra of \BKld.}
}
\end{figure}

\begin{figure}[h]
\center{
\includegraphics[width=0.6\textwidth]{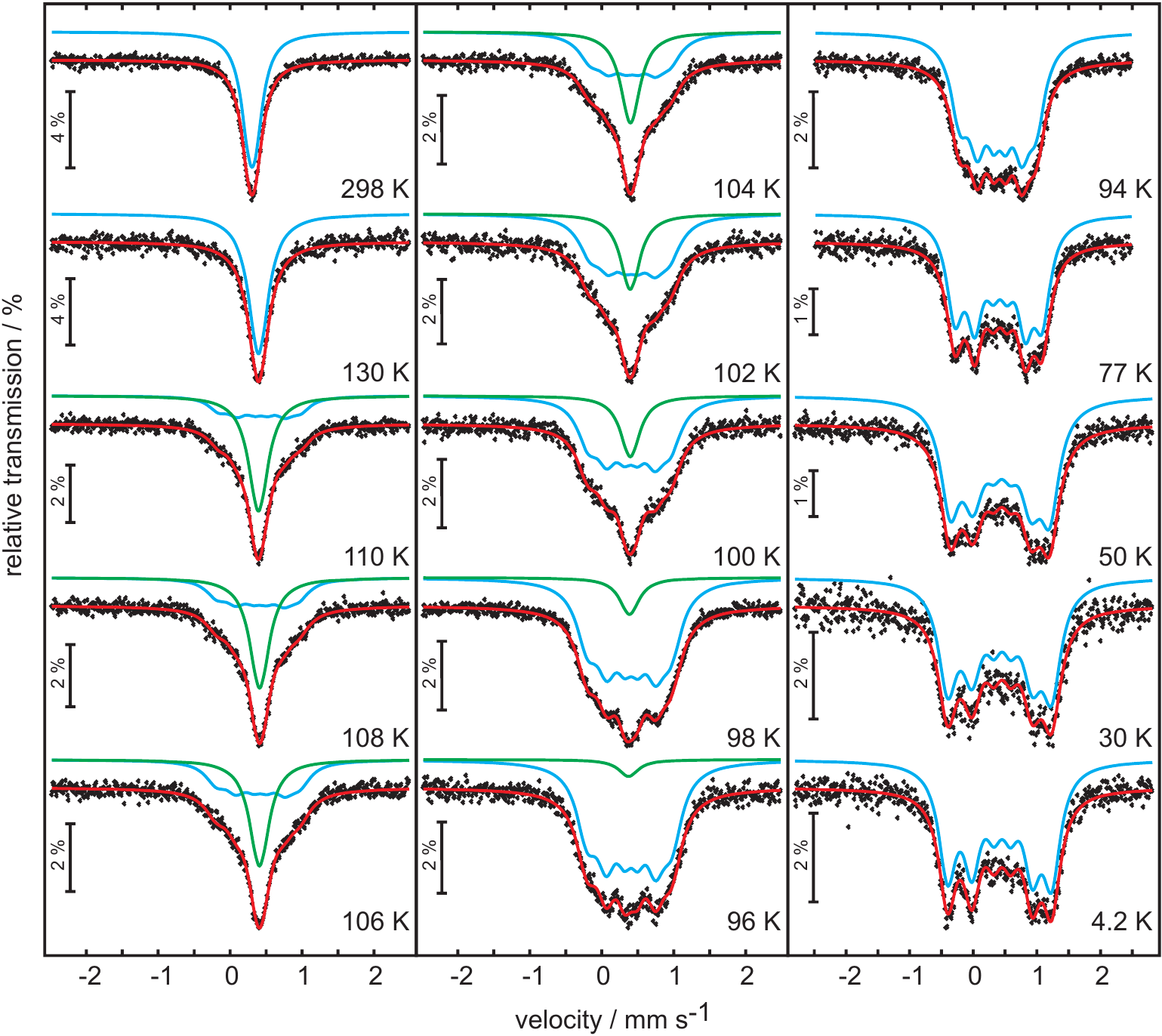}
\caption{\label{fig:MB-BKmd} (Color online) \MB-spectra of \BKmd.}
}
\end{figure}

The potassium-doped samples \BKFA~with $x$ = 0.1 and 0.20 show strikingly different temperature dependencies of the \MB~spectra, shown in Figs.~\ref{fig:MB-BKld} and \ref{fig:MB-BKmd}. In the temperature ranges 148-136 K ($x$ = 0.1) and 110-96 K ($x$ = 0.2), the spectra show superpositions of one magnetically split and one un-split component. The non-magnetic components rapidly diminishes within small temperature ranges ($\approx$ 10 K). Below 136 K and 96 K, respectively, the spectra of \BKld~and \BKmd~can be well reproduced by one single magnetically split signal. These magnetic transition temperatures are very close to the structural transition temperatures extracted from the X-ray data. Similar to pure \BFA, the magnetic hyperfine fields increase with decreasing temperature. \BFA~and \BKld~show almost similar hyperfine fields of $\approx$ 5.5 T at 4.2 K (Table~\ref{tab:MB-Data}), while a decrease ob $B_{hf}$ by 10\% is observed for \BKmd.\\

These findings suggest that the spectra of the underdoped compounds ($x$ = 0.1 and 0.2) are caused by temperature-dependent superpositions of paramagnetic and antiferromagnetically ordered domains. This reflects a chemical inhomogeneity of the Ba/K distribution, where the un-split components represent domains with higher potassium contents and lower N\'eel temperatures and vice versa. This in agreement with the smaller isomer shifts of the un-split signals, which indicate higher doping. On cooling, more domains get magnetically ordered until the paramagnetic fractions are completely consumed. Thus we see no distinct phase separation, but a continuous (narrow) distribution of the potassium concentrations. Since no paramagnetic component exists below $\approx$ 136 K and $\approx$ 96 K, respectively, we observe homogenous co-existence of AF magnetic ordering with superconductivity in \BKFA~at $x$ = 0.1 and 0.2.\\

\begin{figure}[h]
\center{
\includegraphics[width=0.4\textwidth]{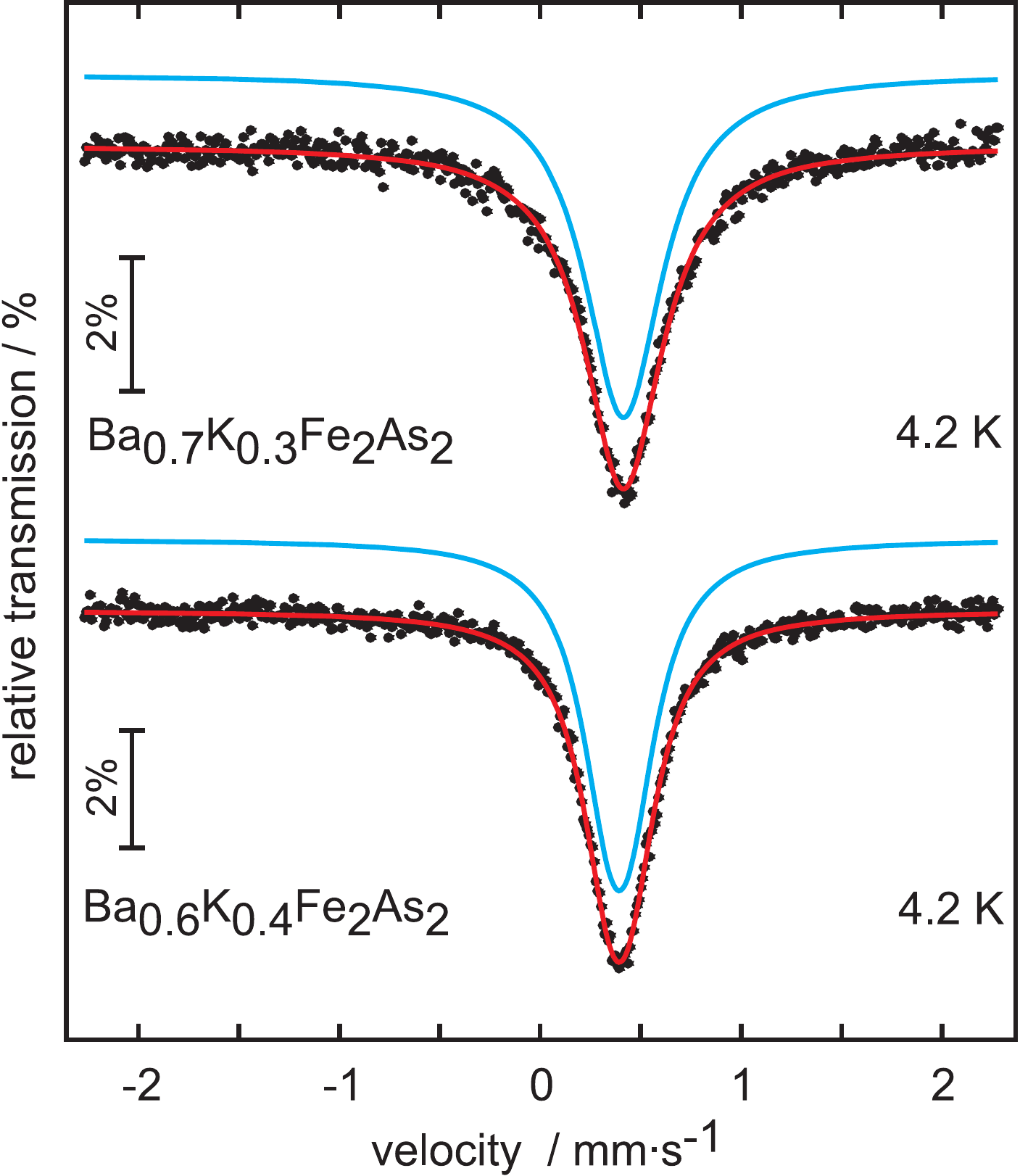}
\caption{\label{fig:MB-BKhd} (Color online) \MB-spectra of \BKhd~and \BKSC~at 4.2 K}
}
\end{figure}

A further increase of doping again radically changes the \MB-spectra. Fig.~\ref{fig:MB-BKhd} shows the spectra of \BKhd~($T_c$ = 33 K) and \BKSC~($T_c$ = 38 K) recorded at 4.2 K together with transmission integral fits. The fitting parameters are listed in Table~\ref{tab:MB-Data-SC}. In both cases we observe only un-split absorption lines, which can be fitted by one component without any magnetic hyperfine field. The line width is slightly increased in comparison with undoped \BFA, hinting at a small inhomogeneity of the Ba/K concentrations. Thus, in contrast to the completely magnetic ordered phase detected at $x$ = 0.2, we find the phases at $x$ = 0.3 and 0.4 to be completely non-magnetic at low temperatures. This is in line with the absence of lattice distortions, since both compounds keep the tetragonal structure at low temperatures (see Fig.~\ref{fig:Lattices} and Ref.~\cite{Rotter-Angewandte})

%

\section{Conclusion}

In summary, we have studied the doping dependencies of the physical properties of \BKFA~in the underdoped region. The SDW anomaly connected with a structural phase transition is continuously suppressed by increasing doping concentrations and no longer observed in \BKhd. Specific heat measurements reveal a smearing of the phase transitions over larger temperature ranges in the underdoped samples, but no SDW anomaly in the optimally doped compounds. This is in agreement with the structural data. Bulk superconductivity was detected in all samples except \BKld. \MB-spectra spectra of the underdoped compounds \BKld~ and \BKmd~ show temperature-dependent superpositions of paramagnetic and antiferromagnetically ordered domains, which reflect chemical inhomogeneities of the Ba/K distributions resulting in different N\'eel temperatures of the domains. At lower temperatures, more domains get magnetically ordered until the paramagnetic fractions are completely consumed. Thus we see no distinct phase separation, but a continuous distribution of the potassium concentrations. No paramagnetic component is observed in \BKmd~below $T_c$ (24 K), suggestive of the co-existence of superconductivity and AF ordering. Only at higher doping levels ($x$ = 0.3), the magnetic and structural phase transitions are completely suppressed and superconductivity reaches the highest $T_c$.

Our results contradict recent reports on mesoscopic phase separations in AF ordered and non-magnetic SC regions in single crystals of almost optimally doped \BKFA~\cite{Aczel-2008,Goko-2008,Keimer-2008}. In the polycrystalline material studied here, the structural distortion and AF ordering are definitely absent already at $x$ = 0.3. The origin of magnetically ordered fractions detected in almost optimally doped single crystals of \BKSC~with high $T_c$ may either be attributed to a strongly inhomogeneous potassium distribution caused by uncontrolled single crystal growth or by magnetic impurity phases like FeAs with a N\'eel temperature of 77 K \cite{FeAs-1972}.

\section{Acknowledgments}
It is a pleasure to thank E.-W.~Scheidt and C. Kant for fruitful discussions. We acknowledge support by the BMBF via contract number VDI/EKM 13N6917 and by the DFG via SFB 484 (Augsburg) and project Jo257/5-1 (M\"unchen).

\section{References}



\providecommand{\newblock}{}

\begin{table}[p]
\caption{\label{tab:Structures} Crystal structure data of Ba$_{1-x}$K$_{x}$Fe$_2$As$_2$ at different temperatures.}

\begin{center}
\begin{tabular}{llllll}
 & \multicolumn{2}{c}{$x=0.1$} & \multicolumn{2}{c}{$x=0.2$} & {$x=0.3$}\\
\hline
 Temp. (K)& 300& 10& 300  & 10& 300\\
 Space group& $I4/mmm$& $Fmmm$& $I4/mmm$  & $Fmmm$& $I4/mmm$\\
 \textit{a} (pm)& 395.37(1)& 560.07(1)& 393.95(1)  & 557.34(1)& 392.57(1)\\
 \textit{b} (pm)& $=a$& 556.20(1)& $=a$  & 554.64(1)& $=a$\\
 \textit{c} (pm)& 1310.60(1)& 1301.35(1)& 1318.90(3)  & 1309.26(3)& 1327.02(3)\\
 \textit{V} (nm$^{3}$)& 0.20487(1)& 0.40538(1)& 0.20469(1)  & 0.40472(1)& 0.20451(1)\\
 \textit{Z}& 2& 4& 2  & 4& 2\\
 data points& 17401& 17401& 17401  & 17401& 17501\\
 reflections& 46& 70& 46  & 70& 46\\
 \textit {d} range &  $1.012 - 6.553$& $1.006 -6.507$&  $1.009 - 6.595$  & $1.005 -  6.546$& $1.007 - 6.635$\\
 R$_{P}$, \textit{w}R$_{P}$& 0.0153, 0.0211& 0.0154, 0.0210& 0.0158, 0.0217  & 0.0172, 0.0229& 0.0146, 0.0192\\
 R$_\textit{bragg}$ , $\chi^2$& 0.0121, 1.319& 0.0132, 1.308& 0.0123, 1.190  & 0.0117, 1.222& 0.0076, 1.212\\
\\  
 K,Ba& 2$a$ (0,0,0)&  2$a$ (0,0,0)& 2$a$ (0,0,0)  & 2$a$ (0,0,0)&2$a$ (0,0,0)\\
& $U_{iso} = 119(4)$& $U_{iso} = 45(4)$& $U_{iso} = 173(7)$  & $U_{iso} = 117(8)$&$U_{iso} = 181(6)$\\
 Fe& 4$d$ ($\frac{1}{2},0,\frac{1}{4}$)&  8$f$ ($\frac{1}{4},\frac{1}{4},\frac{1}{4}$)& 4$d$ ($\frac{1}{2},0,\frac{1}{4}$)  & 8$f$ ($\frac{1}{4},\frac{1}{4},\frac{1}{4}$)&4$d$ ($\frac{1}{2},0,\frac{1}{4}$)\\
& $U_{iso} = 114(4)$& $U_{iso} = 43(4)$& $U_{iso} = 156(7)$  & $U_{iso} = 102(8)$&$U_{iso} = 58(5)$\\
 As& 4$e$ (0,0,$z$)&  8$i$ (0,0,$z$)& 4$e$ (0,0,$z$)  & 8$i$ (0,0,$z$)&4$e$ (0,0,$z$)\\
& $z$ = 0.3547(1)&  $z$ = 0.3538(1)& $z$ = 0.3545(1)  & $z$ = 0.3537(1)&$z$ = 0.3545(1)\\
& $U_{iso} = 147(4)$& $U_{iso} = 57(4)$& $U_{iso} = 129(7)$  & $U_{iso} = 72(8)$&$U_{iso} = 79(5)$\\
 K : Ba ratio& 14(1) : 86(1)& 13(1) : 87(1)& 20(1) : 80(1)  & 20(1) : 80(1)&24(1) : 76(1)\\
\\
Lengths (pm):\\  
Ba--As&  338.3(1)$\times$8& 337.0(1)$\times$4&  338.3(1)$\times$8  & 337.0(1)$\times$4& 337.7(1)$\times$8\\
&& 338.6(1)$\times$4&  & 338.2(1)$\times$4& \\
Fe--As&  240.6(1)$\times$4& 239.1(1)$\times$4&  240.4(1)$\times$4  & 238.9(1)$\times$4& 240.8(1)$\times$4\\
Fe--Fe&  279.6(1)$\times$4& 278.1(1)$\times$2&  278.6(1)$\times$4  & 277.3(1)$\times$2& 277.6(1)$\times$4\\
&& 280.0(1)$\times$2&  & 278.7(1)$\times$2& \\
Angles (deg):\\
As--Fe--As&  110.5(1)$\times$2& 111.2(1)$\times$2&  110.1(1)$\times$2  & 110.7(1)$\times$2& 109.2(1)$\times$2\\
&  109.0(1)$\times$4& 108.9(1)$\times$2&  109.2(1)$\times$4  & 109.0(1)$\times$2& 109.6(1)$\times$4\\
&& 108.3(1)$\times$2&  & 108.6(1)$\times$2& \\
\end{tabular}
\end{center}
\end{table}

\begin{table}[p]
\caption{\label{tab:MB-Data} Fitting parameters of the \MB~spectroscopic measurements of \BFA, Ba$_{0.9}$K$_{0.1}$Fe$_2$As$_2$, Ba$_{0.8}$K$_{0.2}$Fe$_2$As$_2$, and \KFA~at different temperatures. Numbers in parentheses are the statistical errors in the last digit. Values without standard deviations were kept fixed during the fitting procedure. ($\delta$), isomer shift; ($\Gamma$), experimental line width, ($\Delta E_Q$), quadrupole splitting parameter, ($B_{hf}$), magnetic hyperfine field. A1/A2 is the ratio of the signals.}
\begin{center}
\begin{tabular}{lllllllll}
\multicolumn{2}{l}{\textbf{\BFA}} \\
\hline
$T$ & $\delta_1$ & $\Delta E_{Q1}$ & $\Gamma_1$ & $B_{hf}$ & $\delta_2$ & $\Delta E_{Q2}$ & $\Gamma_2$ & A1/A2\\
(K) & (mm s$^{-1}$) & (mm s$^{-1}$) & (mm s$^{-1}$) & (mm s$^{-1}$) & (T) & (mm s$^{-1}$) & (mm s$^{-1}$) & (mm s$^{-1}$)\\
\hline
298 & 0.31(1) &  0.00(1)  & 0.32(1)  &   - \\
155 & 0.40(1) & -0.06(22) & 0.46(1)  & 0.37(58)\\
145 & 0.41(1) & -0.02(1)  & 0.30     & 1.79(9)\\
140 & 0.40(1) & -0.02(1)  & 0.39(4)  & 3.80(7)\\
138 & 0.40(1) & -0.02(1)  & 0.33(3)  & 3.93(4)\\
136 & 0.40(1) & -0.03(1)  & 0.32(2)  & 4.12(3)\\
134 & 0.40(1) & -0.02(1)  & 0.40(4)  & 4.16(3)\\
132 & 0.41(1) & -0.03(1)  & 0.38(3)  & 4.31(3)\\
130 & 0.41(1) & -0.02(1)  & 0.46(6)  & 4.37(2)\\
125 & 0.41(1) & -0.03(1)  & 0.48(6)  & 4.62(2)\\
110 & 0.42(1) & -0.02(1)  & 0.47(7)  & 5.16(1)\\
77  & 0.43(1) & -0.03(1)  & 0.33(2)  & 5.23(1)\\
4.2 & 0.44(1) & -0.04(1)  & 0.25(1)  & 5.47(1)\\
\hline
\end{tabular}
\end{center}
\end{table}

\begin{table}[h]
\begin{center}
\begin{tabular}{lllllllll}
\multicolumn{2}{l}{\textbf{\BKld}}\\
\hline
$T$ & $\delta_1$ & $\Delta E_{Q1}$ & $\Gamma_1$ & $B_{hf}$ & $\delta_2$ & $\Delta E_{Q2}$ & $\Gamma_2$ & A1/A2\\
(K) & (mm s$^{-1}$) & (mm s$^{-1}$) & (mm s$^{-1}$) & (mm s$^{-1}$) & (T) & (mm s$^{-1}$) & (mm s$^{-1}$) & (mm s$^{-1}$)\\
\hline
298  &        &           &         &         & 0.30(1)  &  -0.06(1)  & 0.28(1) \\
160  &        &           &         &         & 0.40(1)  &  -0.04(2)  & 0.32(1) \\
148 & 0.38(1) & -0.06(2)  & 0.32    & 3.32(8) & 0.40(1)  &     -      & 0.30 &  28:72\\
146 & 0.41(1) & -0.02(1)  & 0.32    & 3.46(3) & 0.39(1)  &     -      & 0.30 &  34:66\\
144 & 0.39(1) & -0.02(1)  & 0.32(1) & 3.52(5) & 0.39(1)  &     -      & 0.30 &  46:54\\
142 & 0.40(1) & -0.03(1)  & 0.33(1) & 3.64(2) & 0.40(1)  &     -      & 0.30 &  66:34\\
141 & 0.39(1) & -0.01(1)  & 0.30(1) & 3.61(2) & 0.39(1)  &     -      & 0.30 &  71:29\\
138 & 0.39(1) & -0.03(1)  & 0.39(1) & 3.78(2) & 0.38(1)  &     -      & 0.30 &  91:9\\
136 & 0.40(1) & -0.02(1)  & 0.37(1) & 3.91(3) & 0.38(1)  &     -      & 0.30 &  96:4\\
124 & 0.40(1) & -0.03(1)  & 0.29(1) & 4.58(1)\\
120 & 0.40(1) & -0.03(1)  & 0.28(1) & 4.65(1)\\
118 & 0.41(1) & -0.03(1)  & 0.29(1) & 4.72(1)\\
116 & 0.41(1) & -0.03(1)  & 0.28(1) & 4.77(1)\\
114 & 0.41(1) & -0.03(1)  & 0.28(1) & 4.79(1)\\
112 & 0.41(1) & -0.02(1)  & 0.28(1) & 4.81(1)\\
110 & 0.41(1) & -0.02(1)  & 0.27(1) & 4.86(1)\\
77  & 0.42(1) & -0.03(1)  & 0.27(1) & 5.22(1)\\
50  & 0.44(1) & -0.03(1)  & 0.31(1) & 5.46(2)\\
30  & 0.45(1) & -0.04(1)  & 0.36(1) & 5.55(2)\\
4.2 & 0.44(1) & -0.04(1)  & 0.36(1) & 5.57(2)\\
\hline
\end{tabular}
\end{center}
\end{table}

\begin{table}[h]
\begin{center}
\begin{tabular}{lllllllll}
\multicolumn{2}{l}{\textbf{\BKmd}}\\
\hline
$T$ & $\delta_1$ & $\Delta E_{Q1}$ & $\Gamma_1$ & $B_{hf}$ & $\delta_2$ & $\Delta E_{Q2}$ & $\Gamma_2$ & A1/A2\\
(K) & (mm s$^{-1}$) & (mm s$^{-1}$) & (mm s$^{-1}$) & (mm s$^{-1}$) & (T) & (mm s$^{-1}$) & (mm s$^{-1}$) & (mm s$^{-1}$)\\
\hline
298 &         &           &         &         & 0.30(1) &  -0.09(1) & 0.28(1)\\
130 &         &           &         &         & 0.39(1) &   0.00(1) & 0.34(1)\\
110 & 0.42(1) & -0.02(1) &  0.26(2) & 3.54(7) & 0.39(1) &   -       & 0.33 & 37:63\\
108 & 0.40(1) & -0.03(1) &  0.28(2) & 3.54(6) & 0.41(1) &   -       & 0.33 & 45:55\\
106 & 0.40(1) & -0.03(1) &  0.29(2) & 3.61(6) & 0.41(1) &   -       & 0.33 & 52:48\\
104 & 0.40(1) & -0.02(1) &  0.30(2) & 3.46(5) & 0.40(1) &   -       & 0.33 & 61:39\\
102 & 0.40(1) & -0.02(1) &  0.30(2) & 3.46(4) & 0.40(1) &   -       & 0.33 & 72:28\\
100 & 0.39(1) & -0.02(1) &  0.29(1) & 3.57(3) & 0.40(1) &   -       & 0.33 & 79:21\\
98  & 0.40(1) & -0.02(1) &  0.29(1) & 3.61(1) & 0.39(1) &   -       & 0.33 & 90:10\\
96  & 0.40(1) & -0.02(1) &  0.28(1) & 3.65(2) & 0.37(1) &   -       & 0.33 & 96:4\\
94  & 0.40(1) & -0.02(1) &  0.27(3) & 3.72(4)\\
77  & 0.41(1) & -0.02(1) &  0.29(3) & 4.25(2)\\
50  & 0.43(1) & -0.03(1) &  0.33(4) & 4.85(1)\\
30  & 0.43(1) & -0.04(1) &  0.30    & 5.07(3)\\
4.2 & 0.43(1) & -0.04(1) &  0.31(3) & 5.07(2)\\
\hline
\end{tabular}
\end{center}
\end{table}

\begin{table}[h]
\begin{center}
\begin{tabular}{lllllllll}
\multicolumn{2}{l}{\textbf{\KFA}}\\
\hline
$T$ & $\delta_1$ & $\Delta E_{Q1}$ & $\Gamma_1$ & $B_{hf}$ & $\delta_2$ & $\Delta E_{Q2}$ & $\Gamma_2$ & A1/A2\\
(K) & (mm s$^{-1}$) & (mm s$^{-1}$) & (mm s$^{-1}$) & (mm s$^{-1}$) & (T) & (mm s$^{-1}$) & (mm s$^{-1}$) & (mm s$^{-1}$)\\
\hline
298 & 0.21(1) & -0.03(11) & 0.39(1)\\
77  & 0.33(1) & -0.08(1) & 0.33 \\
4.2 & 0.34(1) & -0.09(4) & 0.40(2)\\
\hline
\end{tabular}
\end{center}
\end{table}

\begin{table}[h]
\caption{\label{tab:MB-Data-SC} Fitting parameters of the \MB~spectroscopic measurements of \BKmd~ and \BKhd~ at 4.2 K.}
\begin{center}
\begin{tabular}{lllll}
\hline
 & $T$ & $\delta_1$ & $\Delta E_{Q1}$ & $\Gamma_1$\\
 & (K)  & (mm s$^{-1}$) & (mm s$^{-1}$) & (mm s$^{-1}$)\\
\hline
\BKhd & 4.2 & 0.41(1) & -0.02*   &  0.47(1)\\
\BKSC & 4.2 & 0.39(1) & -0.10(1) &  0.35(1)\\
\hline
\end{tabular}
\end{center}
\end{table}
\end{document}